\documentclass[sigconf,authorversion,nonacm]{acmart}
\pdfoutput=1
\setcopyright{rightsretained}

\usepackage{amsfonts}
\usepackage{amsmath}
\usepackage{caption}
\usepackage{comment}
\usepackage{subcaption}
\usepackage{graphicx}
\usepackage{pdflscape}
\usepackage{listings}
\usepackage{makecell}
\usepackage{upquote,textcomp}
\usepackage{varwidth}

\usepackage{etoolbox}
\makeatletter
\patchcmd{\maketitle}{\@copyrightspace}{}{}{}
\makeatother

\begin{document}

%\begin{frontmatter}

\title{JSONoid: Monoid-based Enrichment for Configurable and Scalable Data-Driven Schema Discovery}

% %%
% %% The "author" command and its associated commands are used to define the authors and their affiliations.
% \author{Michael J. Mior}
% \affiliation{%
%   \institution{Rochester Institute of Technology}
%   \streetaddress{102 Lomb Memorial Drive}
%   \city{Rochester}
%   \state{New York}
%   \postcode{14619}
% }
% \email{mmior@cs.rit.edu}

\author{Michael J. Mior}
\email{mmior@mail.rit.edu}
\affiliation{
  \institution{Rochester Institute of Technology}
  \streetaddress{102 Lomb Memorial Drive}
  \city{Rochester}
  \state{New York}
  \postcode{14623--5608}
}

%\author{Anonymous Author}

\thispagestyle{plain}
\pagestyle{plain}

%\author[RIT]{Michael J. Mior}
%\ead{mmior@mail.rit.edu}
%\ead[url]{https://cs.rit.edu/~dataunitylab}
%\affiliation[RIT]{organization={Rochester Institute of Technology},
%  addressline={102 Lomb Memorial Drive},
%  city={Rochester},
%  postcode={14623-5608},
%  state={NY},
%  country={USA}}

%\author{\IEEEauthorblockN{Michael J. Mior}
%\IEEEauthorblockA{\textit{Dept. of Computer Science, Rochester Institute of Technology}\\
%Rochester, NY, USA\\
%mmior@mail.rit.edu}
%}

%\maketitle

\begin{abstract}
Schema discovery is an important aspect to working with data in formats such as JSON.
Unlike relational databases, JSON data sets often do not have associated structural information.
Consumers of such datasets are often left to browse through data in an attempt to observe commonalities in structure across documents to construct suitable code for data processing.
However, this process is time-consuming and error-prone.

Existing distributed approaches to mining schemas present a significant usability advantage as they provide useful metadata for large data sources.
However, depending on the data source, ad hoc queries for estimating other properties to help with crafting an efficient data pipeline can be expensive.
We propose JSONoid, a distributed schema discovery process augmented with additional metadata in the form of monoid data structures that are easily maintainable in a distributed setting.
JSONoid subsumes several existing approaches to distributed schema discovery with similar performance.
Our approach also adds significant useful additional information about data values to discovered schemas with linear scalability.

\end{abstract}

\maketitle

%\begin{keyword}
%schema discovery \sep semi-structured data \sep distributed data processing
%\end{keyword}

%\end{frontmatter}

% %%% do not modify the following VLDB block %%
% %%% VLDB block start %%%
% \pagestyle{\vldbpagestyle}
% \begingroup\small\noindent\raggedright\textbf{PVLDB Reference Format:}\\
% \vldbauthors. \vldbtitle. PVLDB, \vldbvolume(\vldbissue): \vldbpages, \vldbyear.\\
% \href{https://doi.org/\vldbdoi}{doi:\vldbdoi}
% \endgroup
% \renewcommand\thefootnote{}
% %%% VLDB block end %%%
%
% %%% do not modify the following VLDB block %%
% %%% VLDB block start %%%
% \ifdefempty{\vldbavailabilityurl}{}{
% \vspace{.3cm}
% \begingroup\small\noindent\raggedright\textbf{PVLDB Artifact Availability:}\\
% The source code, data, and/or other artifacts have been made available at \url{\vldbavailabilityurl}.
% \endgroup
% }
% %%% VLDB block end %%%

\section{Introduction}

Non-relational data formats such as JSON~\cite{Lattak22} have grown significantly in popularity in recent years.
One of the main drivers of this growth is the flexibility provided by requiring little to no upfront schema design.
This has the advantage of accepting a wide variety of data without advance planning.
Such flexibility is useful in domains such as logging where different events with different attributes may be added regularly, or Web services using dynamic languages such as JavaScript and Python are well-suited to ad hoc data processing.

However, this flexibility comes at the cost of providing minimal information to data analysts when consuming the data.
When working with relational data, an analyst can look to the relational schema to provide information on the available data values and their types.
With non-relational data, such a schema is often unavailable, and analysts are left to try to understand the data by manually examining either sample documents or existing source code which processes the data.

In the case of JSON data, JSON Schema~\cite{Bourhis17,Pezoa16} provides a standard mechanism to represent the structure in a collection of JSON documents.
There are many tools built around JSON Schema such as validators and code generators, making it easier to work with data which have an available schema.
While JSON Schemas can be useful, schemas for collections of documents are also not commonly available.
For example, many Web services making use of JSON data only provide written documentation, and many users of document databases which store JSON data do not provide schemas.

Several mechanisms for the discovery of JSON Schemas have been proposed to automate the creation of a valid JSON Schema based on collections of documents, which we summarize in Section~\ref{sec:related_work}.
This allows data analysts to explore schemas in a similar way to those of relational databases.

Here, we focus on an approach that makes use of Apache Spark~\cite{Zaharia16} to perform distributed schema inference.
This enables schema discovery to be performed in a distributed fashion across a large collection of JSON documents.
The key idea behind the approach is to construct a schema that precisely matches each individual document and then combines these schemas recursively to produce a schema that fits the whole collection.
The data structures used in our schema discovery process are all structured in the form of monoids.
Monoids are algebraic structures with an identity element (representing an empty collection) and an associated binary operation (merging schema information).
Restricting the data structures we use to monoids means that we can efficiently maintain all necessary data structures to produce a final JSON Schema in a distributed fashion since individual schemas can be constructed in parallel before merging.

A key element lacking from JSON Schema is information about data size, distributions, and relationships.
In a relational database, this information can often be gleaned from running ad hoc SQL queries.
However, sources of JSON data rarely provide simple facilities for issuing such queries.
We seek to augment the distributed schema discovery process by providing additional useful information about the data while adding minimal overhead in order to maintain scalability.

Our contributions are as follows:
\begin{itemize}
  \item A distributed mechanism for JSON schema discovery that incorporates not only structure, but also summary information from individual data values.
  \item An analysis of schema information content and runtime that incorporates the value of schema enhancements.
  \item A description of a number of use cases for such an enhanced schema, including exploration, constraint discovery, and outlier detection.
\end{itemize}

The remainder of the paper starts by providing some background in Section~\ref{sec:background}.
Next, we provide an overview of our schema discovery process in Section~\ref{sec:discovery}.
We then describe each of our enhancement monoids in Section~\ref{sec:monoids}.
We evaluate the information content of our generated schemas, as well as the runtime performance, in Section~\ref{sec:eval}.
Section~\ref{sec:use_cases} outlines several new use cases enabled by the enhanced schemas created by JSONoid.
We then describe some related work on which JSONoid is based in Section~\ref{sec:related_work}.
Finally, we outline future work and conclude in Sections~\ref{sec:future_work} and~\ref{sec:conclusion}.

\section{Background}\label{sec:background}

\begin{figure}
\centering
\begin{varwidth}{\linewidth}
{\small
\begin{verbatim}
pair: STRING ':' value;
obj: '{' pair (',' pair)* '}' | '{}';
array: '[' value (',' value)* ']' | '[]';
value: STRING | NUMBER | obj | array
      | 'true' | 'false' | 'null';
\end{verbatim}
}
\end{varwidth}
\caption{Simplified JSON grammar~\protect\cite{Parr13}}~\label{fig:json_grammar}
\end{figure}

\begin{figure}
  \begin{subfigure}{0.65\textwidth}
  {\small
  \begin{verbatim}
  {"type": "object",
   "properties": {
     "firstName": { "type": "string" },
     "lastName": { "type": "string" },
     "age": { "type": "integer",
              "minimum": 0 }}}
  \end{verbatim}}
  \vspace*{-6mm}
  \caption{Schema}
  \end{subfigure}
  \begin{subfigure}{0.30\textwidth}
  \vspace*{3mm}
  {\small
  \begin{verbatim}
  {"firstName": "Carla",
   "lastName": "Singh",
   "age": "43"}
  \end{verbatim}
  }
  \vspace*{4.5mm}
  \caption{Document}
  \end{subfigure}

\caption{JSON Schema example~\cite{JsonSchema}}\label{fig:json_schema}
\end{figure}

JavaScript Object Notation (JSON) arose from the JavaScript programming language representation of object literals.
Primitive values are strings, numbers, arrays, booleans, or null.
Objects and arrays can be nested to any depth.
A simplified grammar of JSON is given in Figure~\ref{fig:json_grammar}.
The simplicity and flexibility of JSON has resulted in it becoming a common data format for both Web APIs and document databases.
As mentioned previously, unlike a relational database, it is generally not necessary to provide any schema information in order to store and process JSON data using commonly available document databases and JSON processing libraries.

\subsection{JSON Schema}

This lack of schema information means that developers often must resort to manually examining documents in a collection to gain an understanding of their structure.
This can result in a ``guess-and-check'' approach to development where assumptions are made about the data, and those assumptions are either validated by correctly processing the dataset or broken when another document fails to meet these expectations.
JSON Schema~\cite{Bourhis17,Pezoa16} provides a format for representing the structure of JSON documents, including the attributes that a document should contain and their associated types.
An example of a JSON Schema and a document that is validated against this schema is given in Figure~\ref{fig:json_schema}.
Unfortunately, a JSON Schema describing a dataset is rarely available.
Alternatively, a suitable schema can be mined from a collection of documents.

While these mined schemas are useful, they often fail to provide a complete understanding of the dataset.
If an attribute is optional, what fraction of documents contain this attribute?
Are the values of this attribute unique across the dataset?
How many distinct values exist across the collection?
These questions cannot be answered with purely structural information.
Instead, we must rely on analyzing the data.
JSON Schema is capable of describing only limited information on the data contained in schemas such as minimum and maximum numerical values and regular expressions that strings must match.
Our approach augments this information with additional insight in the generated schemas.
Since the schema discovery process is data-driven, we decide to piggyback on this process to allow capturing of this additional information with minimal overhead since a separate pass over the data is not needed.

There are many existing tools which make the existence of a complete and accurate JSON Schema description of a dataset extremely useful for developers.
For example, libraries exist for most popular languages to generate code for parsing, validation, and data entry.
Schemas can also be used to generate the skeleton of documentation which can be useful for APIs which produce or consume JSON data and need to provide more information on their usage.

\subsection{Apache Spark}

Apache Spark~\cite{Zaharia16} is a computational framework for distributed big data processing.
Spark is similar to MapReduce~\cite{Dean08} in that datasets are loaded from a distributed file system and multiple partitions of a file are processed in parallel.
While Spark does offer a richer computational model than MapReduce, a map-reduce style approach is sufficient to define the semantics of the operations required for JSONoid.
We provide further details on our approach in Section~\ref{sec:discovery}
We use a \texttt{map} function to assign a schema to each individual document in a collection of JSON documents and 
to produce the final schema, we define a \texttt{reduce} function which combines multiple schemas to produce a single schema.
A similar approach is taken by Baazizi et al.~\cite{Baazizi17} although their approach purely collects structural information.
As we describe later, JSONoid mines much richer information which incorporates data values and not merely structure.

%We note that for particularly large schemas, it is possible that a pure map-reduce style approach to execution may be too memory intensive.
%This is because Spark first performs the map stage across each document and then the reduce stage is performed on an entire partition of mapped documents at once.
%Finally, all partitions are then combined together using the reduction operation.
%For especially large collections of documents, the number of partitions may make this reduction prohibitive.
%Since the order of these reductions is unimportant in our monoid-based approach allows for an alternative solution using \emph{tree reduction}.
%In addition to the \texttt{reduce} operator, Spark provides the \texttt{treeReduce} operator.
%This is similar to the \texttt{reduce} operator except that partitions may be merged in multiple stages, reducing the overall size of intermediate results.
%Specifically, a the \emph{level} of the tree may be set such that the number of rounds of reduction is logarithmic in the number of levels.
%This results in significantly smaller intermediate results since the mapper output for partitions are reduced in multiple stages.

\subsection{Probabilistic Data Structures}

Probabilistic data structures (PDS) have become common in big data scenarios, as they allow efficient computation across large datasets without the need for significant memory usage at the expense of accuracy~\cite{Singh20}.
However, many probabilistic data structures have very low levels of error with very significant savings in space.
For the purposes of this work, we make use of probabilistic data structures to solve two problems: approximate membership querying and cardinality estimation.
We describe the specific data structures we use below, with their use detailed in Section~\ref{sec:monoids}.

Approximate membership queries (AMQ) aim to indicate whether a value is a member of a large set with constant query time, while also using a constant amount of space which is expected to be small relative to the size of the set.
For our purposes, this is useful for a succinct summary of the values stored at a particular key in the schema.
AMQ data structures are typically defined such that they cannot return false negatives but may return false positives.
That is, with low probability, a value may be reported as being a member of the set when that is not, in fact, the case.

One of the most common AMQ data structures is the Bloom filter~\cite{Bloom70}.
The simplest version of a Bloom filter uses a bit array to represent the set.
When an item is added, multiple hash functions are computed on the item, and the bits in those positions are set.
To test if an item is in the set, it suffices to check if the appropriate combination of bits is set in the array.
One advantage of Bloom filters in our setting is that they can easily be treated as monoids, as we discuss in Section~\ref{subsec:pds_monoids}.
Furthermore, we can also compare two Bloom filters to see if the set represented by one filter is likely to be a subset of the set represented by another filter.
This is useful for constraint discovery, as we discuss in Section~\ref{subsec:constraints}.

For cardinality estimation, we make use of the HyperLogLog (HLL)~\cite{Flajolet07} data structure.
Cardinality estimation is useful to track the number of unique values that are likely to exist at a particular location within a JSON document.
HLL works similarly to Bloom filters in that it uses hashing of input values to update arrays, which are used to estimate the final value.
Like Bloom filters, we can efficiently merge two HLL data structures, making it easy to treat them as monoids.

\section{Schema Discovery}\label{sec:discovery}

\begin{figure*}
    \centering
    \includegraphics[scale=0.8]{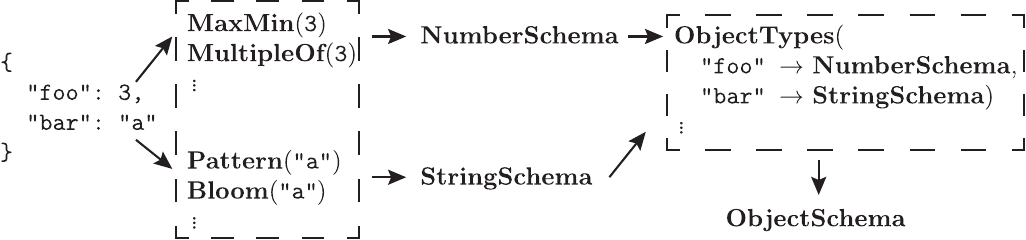}
    \caption{Schema construction for a single document (monoids shown in dashed boxes)}\label{fig:single}
\end{figure*}

\begin{figure*}
    \begin{subfigure}{\textwidth}
    \centering
    \includegraphics[scale=1]{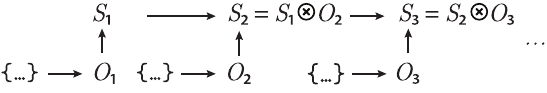}
    \caption{Streaming}\label{fig:streaming}
    \end{subfigure}
    \smallskip\\{\small $O_i$ represents schemas extracted from individual documents while $S_i$ represents the schema on the documents that have been discovered so far}
    \\~\\
    \begin{subfigure}{\textwidth}
    \centering
    \includegraphics[scale=1]{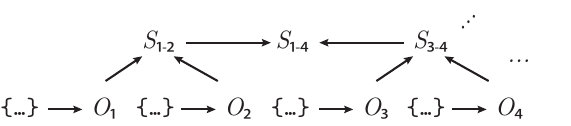}
    \caption{Distributed}\label{fig:distributed}
    \end{subfigure}
    \smallskip\\{\small $O_i$ represents the schema discovered from a single document while $S_{i-j}$ is the combined schema representing a range of documents}
    \caption{Modes of JSONoid schema discovery}\label{fig:strategies}
\end{figure*}

Our goal in the schema discovery is given a collection $\mathbb{J}$ of JSON documents, produce a schema $\mathbf{S}$ such that $\mathbf{S}$ accepts all JSON documents $j\in\mathbb{J}$ as valid.
To avoid the trivial case of a schema that accepts all documents, a good schema should also reject some documents not in $j$.
We say that a schema is more \emph{descriptive} if it is more likely to reject documents which are not in $\mathbb{J}$.
Descriptive schemas are useful for precisely describing a dataset.

Another common use case for JSON schemas is checking whether a document is valid according to the underlying business rules described by the schema.
In this case, an overly descriptive schema may reject documents that should otherwise be valid.
For example, suppose that a numerical value should range from 0 to 100.
A schema discovery algorithm that is highly descriptive will observe a collection of documents and produce a schema that shows the value in the range of 3 to 95 because those were the minimum and maximum values actually observed during the discovery process.
Depending on the specific use case, we may want a schema to be more or less descriptive. 

To this end, schemas in JSONoid consist of a basic type and a configurable set of associated enhancement monoids.
The basic types in JSONoid are object, array, boolean, string, number, or null.
Any additional information about the type (e.g.\ the attributes in an object) is stored in monoids associated with the schema.
These monoids along with their additional information are included based on the desired properties of the final schema.
For primitive values such as strings, the schema can be constructed by building the initial value of all monoids, as discussed in the following section.
For complex values (i.e.\ objects and arrays), schemas are constructed recursively, as shown in Figure~\ref{fig:single}.
Once the schema for a single document has been constructed, the individual schemas for each document in the collection are merged to produce a final schema for the entire collection.

When merging schemas, we evaluate whether they should be combined according to a configurable \emph{equivalence relation}~\cite{Baazizi17}.
Equivalence relations are explained in further detail in Section~\ref{subsec:structure}.
For now, it is sufficient to note that an equivalence relation is a choice made by the user on the desired granularity of the generated schema.
There are two possible cases to consider:

\begin{enumerate}
    \item The two schemas are of the same basic type and are equivalent according to the configured equivalence relation.
    \item The two schemas are different types, or different according to the configured equivalence relation.
\end{enumerate}

In the first case, we merge two schemas by merging their associated monoids, as we discuss in the following section.
In the second case, we create a \emph{product} schema which specifies a schema that can be one of two possible types.
This is equivalent to \texttt{oneOf} in JSON Schema.

We envision two possible merge strategies of operation for our schema discovery process, which are shown in Figure~\ref{fig:strategies}.
The \emph{streaming} strategy is useful if documents appear one at a time in real time, as is the case with data logging.
We start by producing a schema object from the first document, $O_1$, which forms the initial schema of the collection $S_1$.
As new documents appear, we extract their schema and merge it with our schema $S_i$, which contains a schema representing all the documents observed in the stream so far, producing an incrementally more descriptive schema.
Note that in this case we only need to hold a single document in memory at a time and two copies of the schema.
One which is generated from the current document and the incrementally updated schema.
Therefore, the memory used during the discovery process is bounded by a constant factor of the size of the largest document observed.
The runtime of the streaming approach is linear in the number of documents.
We are not aware of existing approaches which implement a similar streaming mode of schema discovery.

Our second merging strategy is \emph{distributed}.
The distributed merge strategy targets large collections of JSON documents stored in distributed file systems.
In distributed mode, construction of the schemas for individual documents can occur in parallel as they are independent of the schema generated for any other document.
We can achieve another level of parallelism in the merge process by organizing schemas into a tree and merging recursively, for example, using \texttt{treeReduce} in Spark.
Both the memory consumption and the runtime of the distributed approach depend on the degree of parallelism.
However, in general, we expect memory consumption to be linear in the degree of parallelism.
The distributed approach therefore consumes more memory than the streaming approach, but is able to produce a schema in logarithmic time relative to the number of documents.
This makes the distributed approach more attractive if all documents are available ahead of time and memory is not constrained.

\begin{comment}
\subsection{Extended Schema Vocabulary}\label{subsec:vocab}

JSON Schema was primarily designed to define a required structure for JSON documents for the purpose of validation.
As such, it only contains data-related constraints that are useful to validate data within a single instance (e.g.\ upper and lower bounds for integer values).
However, since our goal is to provide useful metadata about a collection of JSON documents, we extend this vocabulary to include descriptive features of document collections as a whole.
\end{comment}

\section{Monoids}\label{sec:monoids}

As stated previously, to enhance the discovery process, we design all our enhancements in terms of \emph{monoids}.
In our setting, we use the monoid identity element to initialize the necessary data structures when starting to construct a schema.
The binary operation is used during merging to combine the enhanced data structures of two separate schemas.
We provide detailed descriptions of the various enhancement monoids used in JSONoid in the following subsections.
Each monoid consists of an initial state ($\mathbf{M}_0$) that is constructed from a single value when a new instance of the monoid is required.
Each monoid definition also requires a commutative and associative merge function ($m_1\bigotimes m_2$) that combines the data contained in two monoid instances.

\subsection{Structure Inference}\label{subsec:structure}

We take a similar approach to previous work and recursively construct a schema for each individual attribute of a document.
The result is a schema that perfectly describes the structure of each document.
Once a schema is constructed for each document, we iteratively merge these schemas to produce a final schema that represents the entire collection.
Note that, for scalability, we can produce the individual schemas for each document in parallel.
When merging schemas, we can do so in a tree structure so that $O(log(n))$ merges are required instead of $O(n)$ (where $n$ is the number of documents).

Our approach also incorporates useful elements of past work in schema inference.
Firstly, we consider \emph{parametric} schema inference~\cite{Baazizi19}.
Parametric schema inferences merges objects according to a configurable \emph{equivalence relation}.
If two schemas are equivalent according to the equivalence relation, they are merged.
Otherwise, two separate possible schemas are maintained in a product schema.
This allows the algorithm to trade off between the precision and size of the generated schema.
We incorporate equivalence relations into our approach when merging monoids that represent structural schema information.
Since all monoids are merged (or not) according to the equivalence relation, this gives an easy method to adapt the granularity of the mined schema.

Specifically, JSONoid currently supports both \emph{kind} and \emph{label} equivalence.
Kind equivalence means that schemas which are of the same kind (e.g.\ both objects) will be merged while label equivalence requires the set of keys which are present in an object to be the same in order to allow merging.
The kind equivalence relation minimizes schema size, while label equivalence maximizes precision.
Although we have only implemented these two equivalence relations, others can be easily added without affecting the rest of the discovery process.

Our approach to structure inference also works well with counting types~\cite{Baazizi17b} that associate observed counts with various elements of the schema.
For example, a counting type for an object would indicate the number of times the object was observed, as well as the number of times each attribute in the object was observed.
This is important because objects in JSON document collections do not need to contain the same attributes.
This is helpful for augmenting the schema with additional information such as whether an attribute is expected to be required or optional.
(That is, we consider an attribute as required if every observed object contains that attribute.)
In addition, within objects, we maintain information about the co-occurrence of attributes.
This is helpful in identifying groups of related attributes, which we describe more in Section~\ref{subsec:structure_annotations}

\subsubsection{ObjectTypes}

Our first monoid, \textbf{ObjectTypes}, tracks the type of value contained with each attribute.
The initial value of this monoid constructed from a single object is a map from attribute keys to the schema of each attribute (discovered recursively as described above).
The merge function takes the union of attribute schemas occurring in only one of the two input monoids and merges the schema of those occurring in both.

\begin{align*}
\mathbf{ObjectTypes}_0: &\;obj \rightarrow \nonumber\\
  & \{types: \{key: schema(value) \nonumber\\
  & \mathrm{for}\;(key, value)\;\mathrm{in}\;obj\}\} \\
o_1 \bigotimes o_2 = & \{key: o_1(key)\mathrm{for}\;key\;\mathrm{in}\nonumber\\
  & o_1.types.keys\setminus o_2.types.keys\}\bigcup\nonumber\\
  & \{key: o_2(key)\mathrm{for}\;key\;\mathrm{in}\nonumber\\
  & o_2.types.keys\setminus o_1.types.keys\}\bigcup\nonumber\\
  & \{key: o_1(key)\otimes o_2(key)\mathrm{for}\;key\;\mathrm{in}\nonumber\\
  & o_1.types.keys\cap o_2.types.keys\} \nonumber\\
\end{align*}

\subsubsection{ArrayType}

The next monoid, \textbf{ArrayType} has the same purpose for arrays as \textbf{ObjectType} does for objects, to track the type of values contained within the array.
Here, we consider arrays where all items have the same schema.
(Note that this does not mean that all items must have the same type.
The schema could be a product schema as described above, which accepts multiple types.)
We construct the initial type contained in the monoid by reductive merging of all the types in the array.
On merging, we merge the two schemas of each type.
This allows the construction of arrays where elements may have different types.

\begin{align*}
\mathbf{ArrayType}_0: &\;arr \rightarrow \{type: \mathrm{foldr}(\otimes, arr)\} \nonumber\\
a_1 \bigotimes a_2 = & \{type: a_1.arr\otimes a_2.arr\}
\end{align*}

We note that there is a special case to be considered for the \textbf{ArrayType} schema.
This is the case where all arrays at that particular path in the schema are of fixed identical length.
JSON Schema documentation refers to this as \emph{tuple validation}.
When validating tuples, each element in the tuple may have a distinct type.
This may be used, for example, when representing tabular data in JSON format.
One possible representation is to have nested arrays of rows where each row represents a list of column values, each of which contains the same type.
This is common when tabular data are represented in JSON format, which is common in open data settings~\cite{Moller21}.

The details are not included in the monoid definition above, but to support this use case, the implementation proceeds by maintaining an array of separate schemas, one for each element of the tuple.
This continues on successive merge processes as long as the observed arrays are the same length.
As soon as two monoids containing arrays of separate length are merged, the monoid collapses to the case of \emph{array validation} as presented above.

\subsection{Data Sampling}

One useful view of data values is to provide specific examples of values that occur in the data set.
Depending on the source of the JSON data, viewing sample column values is rarely as simple as in SQL, e.g.\ \texttt{SELECT column FROM table LIMIT 10}.
For example, JSON documents can be streamed as log entries or provided via calls to a Web service.
Furthermore, all documents may not have all possible attributes since some may be optional.
Having a sample set of data values when viewing a schema is helpful to understand what values must be handled when writing data processing code.

For each primitive data type, we collect a sample of the values available across all documents.
A common technique for sampling large collections of elements is reservoir sampling~\cite{Li94}.
The reservoir sampling process uses the probability that each element will be included in a sample to decide whether a newly observed element should be included, evicted from a previous sample, or discarded.
Reservoir sampling is easily adopted in a distributed setting by tracking the total data size in each partition and combining two samples by repeated weighted sampling based on the sizes of the underlying data from which each of the two samples was drawn.
Our monoid that retains a randomized list of examples is described below.
\begin{align*}
\mathbf{Examples}_0: &\;value \rightarrow \{examples: [value], total: 1\} \nonumber\\
e_1 \bigotimes e_2 = & \{examples: \textrm{sample}(\nonumber\\
  & \quad e_1.examples, e_1.total, \nonumber\\
  & \quad e_2.examples, e_2.total), \nonumber\\
  & total: e_1.total + e_2.total\}
\end{align*}

Initialization of the monoid takes the single input value and wraps it in an array.
The merge function considers the second array to be a continuation of the examples presented to the reservoir sampling algorithm.
We also record the total number of values used to generate the sample.
This is used to weight the samples taken from each monoid.
If monoid $e_1$ was constructed by observing 10 values, while monoid $e_2$ was constructed by observing 100 values, we want a 10$\times$ higher probability of selecting an example from $e_2$.

\subsection{Probabilistic Data Structures}\label{subsec:pds_monoids}

For numeric values, we provide a histogram that estimates the distribution of the observed values.
For this we use a streaming histogram proposed by Ben-Haim and Tom-Tov~\cite{Ben10}.
This algorithm maintains a set of $\left(value, count\right)$ pairs which are buckets in the histogram.
New samples either increment the count of an existing bucket with a matching value or add a new bucket with a count of one.
When a new sample causes a configurable maximum number of buckets to be exceeded, the buckets with the closest values are merged by adding their counts and taking a weighted average of their values.
To merge histograms, they are concatenated and sorted by value and bins are merged to return to the original number of bins using the same merge process as for new samples.
The histograms produced in the schema are useful to suggest to developers what range of values to expect.

We also make use of HyperLogLog (HLL)~\cite{Flajolet07} to estimate the number of distinct elements in a set.
HLL maintains a set of registers which are incremented based on a hash function applied to observed values.
These registers are later used to estimate the count of distinct elements.
Since updates to these registers always store the maximum of the current and proposed values, two HLL structures can be merged by taking the maximum values of the registers in the two structures.
The goal of the HLL monoid is to provide a better understanding of the cardinality of values across the dataset.

Finally, we use Bloom filters~\cite{Bloom70}, which use bit arrays and a series of hash functions to support approximate membership queries.
Specifically, using space much smaller than the size of the set of values, Bloom filters can report whether a set contains an item with no false negatives and a configurable probability of false positives.
To combine filters, we take the bitwise OR of two filters.
We construct a Bloom filter monoid for string and numeric values.

Both the HLL and Bloom filter PDS monoids have the same structure for their definition as is given below.
The initial value is added to a new data structure, and when merging values, we use the merge operator of the corresponding data structure ($\otimes$).
For HLL, this is taking the maximum of registers and for Bloom filters, the bitwise OR of the bit vector.
As we describe in Section~\ref{subsec:constraints}, these monoids are also useful for discovering constraints within the data.
\begin{align*}
\mathbf{PDS}_0: &\;value \rightarrow \{pds: \mathrm{PDS}(value)\} \nonumber\\
v_1 \bigotimes v_2 = & \{pds: o_1.pds \otimes o_2.pds\}
\end{align*}

\subsection{Statistics}\label{subsec:stats_monoids}

For numerical measures, in addition to sampling values, it is helpful to provide statistical measures calculated on these values.
Once again, we want to be able to calculate these measures in a distributed fashion, requiring the ability to merge values calculated across multiple partitions.
For example, this can be done for the mean by maintaining a count and sum for each partition and summing values when combining partitions.
The mean can then be calculated in the standard manner by dividing the sum by the count.

\begin{align*}
\mathbf{Mean}_0: &\;value \rightarrow \{sum: value, count: 1\} \nonumber\\
n_1 \bigotimes n_2 = & \{sum: o_1.sum + o_2.sum \nonumber\\
  & \quad total: n_1.total, n_2.total \}
\end{align*}

Other statistics require more complicated calculations, but there has been significant work in calculating common statistical measures in an online setting.
For further statistics, Knuth~\cite{Knuth14} provides an algorithm for calculating the standard deviation based on its corresponding recurrence relation.
We adopt these online calculations of statistical measures as monoids, including others for skewness and kurtosis described by Cook~\cite{Cook14}.
This can be useful both for data understanding and when writing data processing code to select appropriate data structures and estimate their memory usage.
We do not provide the full details of the monoid here, but we defer to the work referenced above.

\subsection{Structural Annotations}\label{subsec:structure_annotations}

An additional monoid we include applies to object types within the JSON data.
We keep track of the counts each time a key is observed within an object, and these counts are recursively accumulated for each separate object observed.
The total number of objects observed is also monitored.
Then the percentage of objects with each given key can be presented as with prior work on counting types~\cite{Baazizi17b}.
\begin{align*}
\mathbf{AttributeCounts}_0: &\;obj \rightarrow  \{counts: \nonumber\\
  & \quad \{key: 1\;\mathrm{for}\;key\;\mathrm{in}\;obj.keys\}\} \\
o_1 \bigotimes o_2 = & \{key: o_1(key)+o_2(key)\mathrm{for}\;key\;\mathrm{in}\nonumber\\
  & o_1.counts.keys\cap o_2.counts.keys\} \nonumber \\
o_i(key) = &\;0\;\mathrm{if}\;key\;\not\in o_i.keys
\end{align*}

This monoid gives the data consumer knowledge of how likely a particular key is to occur.
This lets an analyst know whether they can generally rely on an attribute being present or not.
JSON Schema does not provide such information.
To produce similar information that can be used within the JSON Schema standard, we also define a monoid that populates the \texttt{required} property.
The \texttt{required} property specifies which attributes are present in all instances of an object.
\begin{align*}
\mathbf{Required}_0: &\;obj \rightarrow  \{keys: obj.keys\} \nonumber \\
o_1 \bigotimes o_2 = & \{key: o_1(key)\cap o_2(key)\mathrm{for}\;key\;\mathrm{in}\nonumber\\
  & o_1.keys.keys\cap o_2.keys.keys\} \nonumber\\
o_i(key) = &\;\emptyset\;\mathrm{if}\;key\;\not\in o_i.keys
\end{align*}

Another interesting example of a useful monoid is one used to track whether an array contains unique elements.
Here, we make use of the \textbf{Examples} monoid, which tracks unique examples for a schema.
Since the schema and associated monoids are constructed bottom-up, we have available the examples of array elements at the time array monoids are constructed.
We check if the number of (unique) examples is equal to the length of the array.
If this is the case, each array element is unique.
When merging this monoid, we consider the elements of the combined array to contain unique elements if each array being merged also contains unique elements.
\begin{align*}
\mathbf{Unique}_0: &\;arr \rightarrow  \{isUnique: |arr.examples| == |arr|\} \nonumber \\
a_1 \bigotimes a_2 = & \{isUnique: a_1.isUnique\wedge a_2.isUnique\}
\end{align*}

One final property we maintain for objects is \emph{dependencies}.
A dependency indicates that when one attribute occurs, another set of attributes also occurs.
This can be helpful in identifying possible groups of related attributes.
For example, attributes \texttt{city} and \texttt{state} can both be considered optional. But there may be a dependency such that whenever the attribute \texttt{city} occurs, the attribute \texttt{state} must also be present.
We omit the details here, but dependencies are tracked by counting each time two attributes occur together as well as the total number of occurrences of each attribute individually.
If the number of co-occurrences matches the total count of each attribute, a dependency has been found.

\subsection{Value Restrictions}

Existing schema discovery tools focus primarily on structural and type constraints.
However, little attention has been given to what values within a type are acceptable.
We define several monoids that are used to impose meaningful restrictions on values that are expressed within the data being mined.

\subsubsection{MaxMin}

The first of these value restrictions are minimum and maximum constraints.
This applies to numeric values as well as the length of strings and arrays.
These values can be easily maintained as monoids by starting with the initial value and then taking the minimum (or maximum) of the current value and the monoid being merged.
The result after extracting and merging schemas from all documents is the minimum and maximum value across all documents.
\begin{align*}
\mathbf{MaxMin}_0: &\;num \rightarrow \{min: num, max: num\} \nonumber\\
n_1 \bigotimes n_2 = \{&min: \mathrm{min}(n_1.min, n_2.min), \nonumber\\
  & max: \mathrm{max}(n_1.max, n_2.max)\}
\end{align*}

\subsubsection{Multiple}
There are also additional type-specific restrictions that can be mined.
For example, JSON Schema supports the \texttt{multipleOf} restriction for numerical values, which indicates that a numerical value must be a multiple of a specific constant.
To track this property during schema discovery, we use Euclid's algorithm for computing the greatest common divisor (GCD).
\begin{align*}
\mathbf{Multiple}_0: &\;num \rightarrow \{multiple: num\} \nonumber\\
n_1 \bigotimes n_2 = & \{num: \textrm{gcd}(n_1.multiple, n_2.multiple\}
\end{align*}

The first numerical value encountered is treated as the GCD.
Any subsequent values are compared with this GCD.
Finally, the \texttt{multipleOf} property is emitted if the GCD is greater than 1 to avoid the trivial case where any integer value is accepted.
(In this case, we simply use the type \texttt{integer} instead of \texttt{number}.)

\subsubsection{Pattern}

For string values, one useful property is whether string values match a particular regular expression.
In this work, we consider the simple case where all strings have a common prefix and/or suffix.
We initialize the prefix (and suffix) to the first string value encountered.
On processing subsequent values, we take the longest common prefix (or suffix) of the new value and the previous prefix (or suffix).
\begin{align*}
\mathbf{Pattern}_0: &\;str \rightarrow \{prefix: str, suffix: str\} \nonumber\\
s_1 \bigotimes s_2 = & \{prefix: \nonumber\\
  & \mathrm{common\_prefix}(s_1.prefix, s_2.prefix)\} \nonumber\\
  & suffix: \nonumber\\
  & \mathrm{common\_suffix}(s_1.suffix, s_2.suffix)\}
\end{align*}

The final property is supported in JSON Schema by the \texttt{pattern} property of string values.
Patterns are regular expressions, so we anchor the expression at the start or end of a string or both depending on whether a non-empty prefix, suffix, or both are obtained.
This can handle cases such as a URL that has a common prefix and file extension, providing additional context for anyone making use of the data.
Extending this approach to more complex regular expressions is left for future work.

\subsubsection{Format}

Finally, we also detect possible formatted strings within JSON data.
String formats are commonly used values such as dates (a datatype not natively supported in JSON), GUIDs, and URLs.
To detect formats, we start by assuming that a string property has no defined format.
A format detection function is written for each possible string format to determine if the string values being mined match the given format.
We assume that each string property will only match a single given format, so we check all these functions to see if a string matches any defined formats.
Assuming all values encountered for a string property match a given format, we include this format in the generated schema.
\begin{align*}
\mathbf{Format}_0: &\;str \rightarrow \{format: \mathrm{format}(str)\} \nonumber\\
s_1 \bigotimes s_2 = \{&format: \mathrm{if}\;s_1.format = s_2.format \nonumber\\
  & \mathrm{then}\;s_1.format\;\mathrm{else}\;\emptyset\}
\end{align*}

\section{Evaluation}\label{sec:eval}

Our implementation of JSONoid consists of approximately 9,000 lines of Scala code that produces schemas targeting JSON Schema Draft 2019-09 and is available on GitHub\footnote{\url{https://github.com/dataunitylab/jsonoid-discovery/}}.
We evaluate our implementation on two different dimensions: information content and runtime.
The goal of the evaluation of information content is to identify the value of the additional information provided by our monoids.
The runtime evaluation serves to show the scalability and efficiency of our monoid-based mining approach.
Runtime is important when the goal is to produce a complete description of a large collection of documents that an analyst wishes to examine.
While in some cases this may be done offline, this is prohibitive as the number of collections grows, as is the case in a data lake scenario.
For our evaluation, we make use of the kind equivalence relation in JSONoid.

\subsection{Information Content}

Existing JSON schema inference techniques focus only on extracting the structure of a dataset, with no attention paid to the values each contains.
The primary goal of our schema extraction approach is to provide additional useful information about the datasets under consideration.
To measure the information content of a schema, we start with the simplest possible schema, $S_{min}$, which contains only the monoids \textbf{ObjectType} and \textbf{ArrayType}.
Together these monoids provide basic structural information about a collection of documents, similar to the approach of early work by Baazizi et al.~\cite{Baazizi17}.
We then consider the effectiveness of additional monoids in describing a collection of documents.

In order to do this, we require a ground truth.
Specifically, we require a manually constructed schema which expresses the intent of its author to match against a collection of documents.
We also require a collection of documents that were written in order to adhere to this schema.
For our evaluation, we use \texttt{package.json}, the metadata provided by package developers in the Node.js (JavaScript) ecosystem.

We start with a manually authored JSON Schema, $S_0$ which expresses the structure of \texttt{package.json} files which we obtained via JSON Schema Store~\cite{packageJson}.
To obtain documents that match this schema, we downloaded metadata for the 1,000 top packages from the npm registry.
In order to ensure the integrity of our analysis, we validated these documents against the schema and discovered that 6 of these documents were not valid against the schema and thus excluded from our analysis.
For correctness, we require that the schema which we generate against a collection of documents treats every document in that collection as valid.
Indeed, we find that this is the case regardless of which monoids are used during schema creation.

\begin{figure}
  \begin{subfigure}{0.6\textwidth}
%{\scriptsize
    \begin{verbatim}
{"name": "minami", "version": "1.2.3",
 "dist": {"shasum": "99b6dcdfb2...",
          "tarball": "https://..."}}
    \end{verbatim}
%    }
%    \vspace*{12.5mm}
    \caption{Actual document}
    \vspace*{5mm}
  \end{subfigure}
  \begin{subfigure}{0.4\textwidth}
%{\scriptsize
    \begin{verbatim}
{"os": ["aj8KwZWoIf","n1",
        "eDcvfCpwK9","Z"],
 "version": "t1KSAC",
 "funding": {"url":"BF1gv"},
 "bundleDependencies": ["7aJJei"],
 "directories": {
   "test": "tests",
   "doc": "reHPwJ"}}
 \end{verbatim}
%    }
%    \vspace*{-4mm}
    \caption{Random document}
  \end{subfigure}
    \caption{package.json example}\label{fig:pkg_json}
\end{figure}

To determine the fidelity of our mined schemas to the created schema, we must first have a set of documents which are similar to real-world documents, but different enough that they are not valid according to the schema.
Since our minimal schema consisting only of \textbf{ObjectType} and \textbf{ArrayType} is the simplest possible representation of the document structure, we generate random documents according to this schema.
That is, we generate documents that have attributes with the same names and types as those that exist in the minimal schema.
For each object, we randomly decide which properties to include and then generate values of the correct type for those properties.
This continues recursively for any values which are nested objects.
However, because the minimal schema does not consider constraints on the values within the document, many of these randomly generated documents will be invalid according to the actual schema, $S_0$.

To create our final collection of documents for comparison, we generated random documents conforming to $S_{min}$ as above and then filtered those documents to only those which do \emph{not} conform to $S_0$.
We considered two sources of values for these generated documents: \emph{sampled}, and \emph{random}.
In the sampled case, the values generated for each attribute were selected from the same attribute in our real-world collection of documents.
For the random case, we generated completely random values which have the same type, but may lie outside of the domain of the original documents.
For example, a string may not be in the correct format, such as a URL, or a numeric value may lie outside a valid range.
An example of a real document and a randomly generated document is given in Figure~\ref{fig:pkg_json}.
In this case, the randomly generated document is invalid because the value \texttt{url} is not a valid URL.
We note that this invalid value cannot be detected by simply identifying that values of this attribute must be a string with a particular length, as is done by most existing approaches.
In this case, the \textbf{Format} monoid is needed to identify that the value must be a valid URI.

We also want to identify the possibility that our generated schemas will be overfit to a specific set of observed documents.
In this case, the schema may reject documents that are very similar because specific values within the document have not been previously observed.
This is undesirable in the case where we want to use the generated schema to validate documents or detect outliers.
(But it may be useful to provide a precise summary of a collection.)
To measure how prone our monoids are to overfitting, we split the collected data and created a schema on 90\% of the original dataset followed by validation against the remaining 10\%.
If we avoid overfitting, we would expect the remaining 10\% of the documents to also be valid since they are drawn from the same collection of documents.
Our results are shown in Table~\ref{tbl:accuracy}.
The final column in Table~\ref{tbl:accuracy} represents the fraction of documents that were \emph{not} valid according to the schema generated from the remaining 10\%.
Values closer to 1 represent a higher level of overfitting.
Note that due to excessive runtime, we were only able to generate the schema based on half of the training set for the approach of Spoth et al.
While other approaches completed within seconds, waiting more than 24h did not produce a result.
We believe that this is due to the large number of distinct attribute names present in this dataset.
As such, we expect the performance of this method to be lower than if the schema were generated using the entire training set.

Some monoids are more prone to overfitting than others.
Specifically, the \textbf{MaxMin} monoid applied to string and array length is the most prone to overfitting in our setting.
This is because the test set of documents contains strings of lengths outside the bounds observed when the schema was created.
However, monoids such as \textbf{Required} are less prone to overfitting, since this will only happen when a document \emph{without} has a property that should be considered optional is never observed when creating the schema.
We find that, depending on the selection of monoids, our approach has higher accuracy than any existing approach, with comparable levels of overfitting.

We note the relatively high accuracy and low overfitting of the approach of Spoth et al.
This is mainly due to the design decision to create an ``open'' schema.
That is, their approach generates a schema which specifies that any properties not defined by the schema are valid.
This reduces accuracy, since properties that should not exist are considered valid.
However, this also significantly reduces overfitting, since any properties that were not observed in the sample (which should still be considered valid) are accepted.
In contrast, our approach and that of Baazizi et al.\ generate ``closed'' schemas, where any property not explicitly mentioned in the schema is considered invalid.
To show the impact of this decision, we also modified the schema produced by the approach Spoth et al.\ to produce a closed version.
As expected, the accuracy increases, but the overfitting of this schema also increases significantly.
This is because their approach contains a step that attempts to identify distinct types of record in the schema.
For example, if a schema contains some records with only attributes \texttt{a} and \texttt{b} and some records with only attributes \texttt{c} and \texttt{d}, these would be represented as two mutually exclusive types.
However, this approach can lead to overfitting since not all combinations of attributes will be observed.

It is possible that heuristics applied to the final schema could help reduce overfitting.
Consider the case of maximum and minimum string lengths.
If we observe many strings of length 100 and no strings which are any longer, it is very likely that 100 is the true maximum length which should be allowed for a string.
However, if we see string lengths which are more uniformly distributed and the maximum length of any string is 93, it is much less likely that is is the true expected maximum length of string since constraints are more likely to fall on natural boundaries.
We leave such heuristics for reducing overfitting for future work.

\begin{table}
\centering
{\footnotesize
\begin{tabular}{lllll}
\textbf{Monoid}       & \textbf{Context} & \textbf{\begin{tabular}[c]{@{}l@{}}Accuracy\\ (sampled)\end{tabular}} & \textbf{\begin{tabular}[c]{@{}l@{}}Accuracy\\ (random)\end{tabular}} & \textbf{Overfit} \\ \hline
\textbf{MaxMin}       & Number           & 0.033 & 0.038 & 0.011 \\
\textbf{Multiple}     & Number           & 0.033 & 0.038 & 0.011 \\
\textbf{MaxMin}       & String length    & 0.760 & 0.998 & 0.404 \\
\textbf{Pattern}      & String           & 0.229 & 0.882 & 0.053 \\
\textbf{Format}       & String           & 0.032 & 0.504 & 0.011 \\
\textbf{MaxMin}       & Array length     & 0.760 & 1     & 0.404 \\
\textbf{Unique}       & Array items      & 0.474 & 0.094 & 0.106 \\
\textbf{Required}     & Object           & 0.987 & 0.902 & 0.011 \\
\textbf{Dependencies} & Object           & 0.991 & 0.990 & 0.032 \\ \hline
\textbf{All}          &                  & 1     & 1     & 0.426 \\ \hline
                      & \textbf{Baazizi} & 0.986 & 0.902 & 0.777 \\
                      & \textbf{Spoth\textsuperscript{\textasteriskcentered}}   & 0.837 & 0.752 & 0.053 \\
                      & \textbf{Spoth\textsuperscript{\textasteriskcentered}} (closed) & 1 & 1 & 0.968 \\ \hline
\end{tabular}}

{\scriptsize \smallskip \textasteriskcentered This method could not successfully complete discovery on the full training set.}

\caption{Accuracy and overfit of JSONoid schemas}\label{tbl:accuracy}
\vspace{-3mm}
\end{table}

Finally, we note that while we described many other monoids, many of these do not add information content according to the definition given above.
For example, the statistical monoids such as \textbf{Mean} defined in Section~\ref{subsec:stats_monoids} \emph{do} provide additional information to data consumers.
However, this aggregate statistical information does not allow us to distinguish from documents that adhere to a schema that does not include these monoids.
As discussed in Section~\ref{subsec:outliers}, we believe that these statistical monoids as well as others such as the histogram monoid described in Section~\ref{subsec:pds_monoids} could be useful for outlier detection.
We leave the evaluation of outlier detection using these monoids for future work.

\subsection{Runtime}

All runtime evaluation is performed across a cluster of four machines that contain an eight-core Intel Xeon Silver 4110 @ 2.10GHz with an ADATA SX930 240GB SSD and 16GB RAM.
All four of these machines were used to evaluate distributed mode, while a single machine was used to evaluate streaming mode.

To evaluate the performance of the distributed mode of JSONoid, we compare with the discovery approaches proposed by Spoth et al.~\cite{Spoth2021} and Baazizi et al.~\cite{Baazizi17} using implementations provided by the authors.
All approaches were run using Apache Spark 2.4.8 on top of Hadoop 3.2 with all data stored on HDFS.
For the algorithm used by Baazizi et al\. we used the kind equivalence relation, as we do with JSONoid.
As noted previously, we are also able to produce a schema that contains the same information as the reduced structure identification graph of Klettke et al.~\cite{Klettke15}.
However, since their approach cannot scale beyond main memory, we exclude it from our comparative analysis.
We also evaluate with several different sets of monoids whose values can be computed by JSONoid.
First, \texttt{Minimum} which contains only structural information.
Second, \texttt{Simple} which contains only those properties defined in JSON Schema which are inferred by JSONoid.
Finally, we include all monoids defined in JSONoid as described in Section~\ref{sec:monoids}.
%Results are shown in Figure~\ref{fig:runtime_dist}.

The results for varying percentages of data sampled from ten different datasets are shown in Table~\ref{tbl:runtime_dist}.
The first three columns in each group represent JSONoid for varying configurations of monoids.
Each cell displays both the runtime and the fraction of a test set of 10\% of documents that were valid against the generated schema.
For more details on the datasets used, see Spoth et al.~\cite{Spoth2021}.
(Note that, while data were obtained from the same sources, the specific documents used and samples are not identical, so our results do not perfectly match prior work.)

Note that when discovering a minimal set of monoids, which contains information similar to existing methods, our approach nearly always runs faster than \textsc{Jxplain}. and for smaller data sizes, also Baazizi et al.
As expected, when discovering a larger set of monoids, the runtime performance of our method is less than other methods.
However, as shown in the previous section, we are able to discover significantly more information.
Furthermore, users of JSONoid can select the appropriate tradeoff for their application.
In the cases where a schema must be discovered quickly, using a minimal set of monoids can provide useful structural information.
If more time is available for analysis, using a larger set of monoids can provide more detailed information on the dataset.
As a result, we note that a higher acceptance rate of documents from the test set is not necessarily an indication of a better quality schema.
It may be desirable for some applications to construct a more rigid schema that closely fits the input dataset, while other applications may want to be more accepting.
Our approach makes this choice configurable during the mining process by choosing the appropriate set of monoids.
In some cases, the schema mined by JSONoid does not accept any of the training documents as valid.
As discussed previously, this is the result of some monoids fitting the training data very precisely.
The specific monoids used can be controlled more precisely instead of using the predefined sets in Table~\ref{tbl:runtime_dist} if more control is desired.

\begin{table*}[ht]
{\scriptsize
\begin{tabular}{r|cl|cl|cl|cl|cl|cl|cl|cl|cl|cl|cl|cl|cl|cl|cl|}
\multicolumn{1}{l|}{} & \multicolumn{5}{c|}{1\%} & \multicolumn{5}{c|}{50\%} & \multicolumn{5}{c|}{90\%}\\ \hline
\multicolumn{1}{r|}{\textbf{Dataset}} & \multicolumn{1}{c|}{Min} & \multicolumn{1}{c|}{Simple} & \multicolumn{1}{c|}{All} & \multicolumn{1}{c|}{Spoth} & \multicolumn{1}{c|}{Baazizi} & \multicolumn{1}{c|}{Min} & \multicolumn{1}{c|}{Simple} & \multicolumn{1}{c|}{All} & \multicolumn{1}{c|}{Spoth} & \multicolumn{1}{c|}{Baazizi} & \multicolumn{1}{c|}{Min} & \multicolumn{1}{c|}{Simple} & \multicolumn{1}{c|}{All} & \multicolumn{1}{c|}{Spoth} & \multicolumn{1}{c|}{Baazizi}\\ \hline
\textbf{GitHub} & \multicolumn{1}{c|}{\makecell{32.97\\1.000}} & \multicolumn{1}{c|}{\makecell{133.45\\0.213}} & \multicolumn{1}{c|}{\makecell{273.23\\0.213}} & \multicolumn{1}{c|}{\makecell{42.28\\0.803}} & \multicolumn{1}{c|}{\makecell{\textbf{25.63}\\1.000}} & \multicolumn{1}{c|}{\makecell{298.96\\1.000}} & \multicolumn{1}{c|}{\makecell{5489.03\\0.212}} & \multicolumn{1}{c|}{\makecell{12694.39\\0.212}} & \multicolumn{1}{c|}{\textdagger} & \multicolumn{1}{c|}{\makecell{\textbf{59.66}\\1.000}} & \multicolumn{1}{c|}{\makecell{509.84\\1.000}} & \multicolumn{1}{c|}{\makecell{9957.52\\0.532}} & \multicolumn{1}{c|}{\textdagger} & \multicolumn{1}{c|}{\textdagger} & \multicolumn{1}{c|}{\makecell{\textbf{89.02}\\1.000}} \\\hline
\textbf{NYT} & \multicolumn{1}{c|}{\makecell{\textbf{23.57}\\1.000}} & \multicolumn{1}{c|}{\makecell{29.10\\0.947}} & \multicolumn{1}{c|}{\makecell{35.11\\0.947}} & \multicolumn{1}{c|}{\textdagger} & \multicolumn{1}{c|}{\makecell{24.63\\1.000}} & \multicolumn{1}{c|}{\makecell{34.69\\1.000}} & \multicolumn{1}{c|}{\makecell{150.34\\0.000}} & \multicolumn{1}{c|}{\makecell{350.65\\0.000}} & \multicolumn{1}{c|}{\textdagger} & \multicolumn{1}{c|}{\makecell{\textbf{27.50}\\1.000}} & \multicolumn{1}{c|}{\makecell{41.78\\1.000}} & \multicolumn{1}{c|}{\makecell{251.04\\0.000}} & \multicolumn{1}{c|}{\makecell{602.88\\0.000}} & \multicolumn{1}{c|}{\textdagger} & \multicolumn{1}{c|}{\makecell{\textbf{27.39}\\1.000}} \\\hline
\textbf{Pharma} & \multicolumn{1}{c|}{\makecell{25.25\\0.915}} & \multicolumn{1}{c|}{\makecell{29.83\\0.746}} & \multicolumn{1}{c|}{\makecell{34.93\\0.746}} & \multicolumn{1}{c|}{\textdagger} & \multicolumn{1}{c|}{\makecell{\textbf{24.34}\\1.000}} & \multicolumn{1}{c|}{\makecell{175.84\\0.998}} & \multicolumn{1}{c|}{\makecell{341.53\\0.983}} & \multicolumn{1}{c|}{\makecell{806.49\\0.983}} & \multicolumn{1}{c|}{\textdagger} & \multicolumn{1}{c|}{\makecell{\textbf{29.54}\\1.000}} & \multicolumn{1}{c|}{\makecell{312.95\\0.999}} & \multicolumn{1}{c|}{\makecell{629.55\\0.988}} & \multicolumn{1}{c|}{\makecell{1474.89\\0.988}} & \multicolumn{1}{c|}{\textdagger} & \multicolumn{1}{c|}{\makecell{\textbf{31.71}\\1.000}} \\\hline
\textbf{Twitter} & \multicolumn{1}{c|}{\makecell{\textbf{23.51}\\0.996}} & \multicolumn{1}{c|}{\makecell{33.39\\0.000}} & \multicolumn{1}{c|}{\makecell{40.82\\0.000}} & \multicolumn{1}{c|}{\makecell{25.46\\0.003}} & \multicolumn{1}{c|}{\makecell{25.08\\1.000}} & \multicolumn{1}{c|}{\makecell{43.34\\1.000}} & \multicolumn{1}{c|}{\makecell{344.12\\0.000}} & \multicolumn{1}{c|}{\makecell{742.33\\0.000}} & \multicolumn{1}{c|}{\makecell{59.65\\0.000}} & \multicolumn{1}{c|}{\makecell{\textbf{26.89}\\1.000}} & \multicolumn{1}{c|}{\makecell{59.13\\1.000}} & \multicolumn{1}{c|}{\makecell{677.11\\0.000}} & \multicolumn{1}{c|}{\makecell{1499.44\\0.000}} & \multicolumn{1}{c|}{\makecell{71.73\\0.000}} & \multicolumn{1}{c|}{\makecell{\textbf{28.73}\\1.000}} \\\hline
\textbf{Yelp-Business} & \multicolumn{1}{c|}{\makecell{\textbf{22.85}\\0.999}} & \multicolumn{1}{c|}{\makecell{28.55\\0.977}} & \multicolumn{1}{c|}{\makecell{31.27\\0.977}} & \multicolumn{1}{c|}{\makecell{23.79\\0.788}} & \multicolumn{1}{c|}{\makecell{24.50\\1.000}} & \multicolumn{1}{c|}{\makecell{30.44\\1.000}} & \multicolumn{1}{c|}{\makecell{140.46\\0.151}} & \multicolumn{1}{c|}{\makecell{291.07\\0.151}} & \multicolumn{1}{c|}{\makecell{32.33\\0.788}} & \multicolumn{1}{c|}{\makecell{\textbf{26.35}\\1.000}} & \multicolumn{1}{c|}{\makecell{34.52\\1.000}} & \multicolumn{1}{c|}{\makecell{235.54\\0.151}} & \multicolumn{1}{c|}{\makecell{503.32\\0.151}} & \multicolumn{1}{c|}{\makecell{41.86\\0.788}} & \multicolumn{1}{c|}{\makecell{\textbf{28.67}\\1.000}} \\\hline
\textbf{Yelp-Checkin} & \multicolumn{1}{c|}{\makecell{22.65\\1.000}} & \multicolumn{1}{c|}{\makecell{24.06\\1.000}} & \multicolumn{1}{c|}{\textdagger} & \multicolumn{1}{c|}{\makecell{\textbf{22.61}\\1.000}} & \multicolumn{1}{c|}{\makecell{24.45\\1.000}} & \multicolumn{1}{c|}{\makecell{\textbf{23.76}\\1.000}} & \multicolumn{1}{c|}{\makecell{51.14\\1.000}} & \multicolumn{1}{c|}{\textdagger} & \multicolumn{1}{c|}{\makecell{33.27\\1.000}} & \multicolumn{1}{c|}{\makecell{26.12\\1.000}} & \multicolumn{1}{c|}{\makecell{26.16\\1.000}} & \multicolumn{1}{c|}{\makecell{70.96\\1.000}} & \multicolumn{1}{c|}{\textdagger} & \multicolumn{1}{c|}{\makecell{33.81\\1.000}} & \multicolumn{1}{c|}{\makecell{\textbf{25.73}\\1.000}} \\\hline
\textbf{Yelp-Photos} & \multicolumn{1}{c|}{\makecell{\textbf{22.46}\\1.000}} & \multicolumn{1}{c|}{\makecell{24.25\\1.000}} & \multicolumn{1}{c|}{\textdagger} & \multicolumn{1}{c|}{\makecell{22.56\\1.000}} & \multicolumn{1}{c|}{\makecell{24.21\\1.000}} & \multicolumn{1}{c|}{\makecell{25.28\\1.000}} & \multicolumn{1}{c|}{\makecell{42.34\\1.000}} & \multicolumn{1}{c|}{\textdagger} & \multicolumn{1}{c|}{\makecell{\textbf{23.57}\\1.000}} & \multicolumn{1}{c|}{\makecell{25.00\\1.000}} & \multicolumn{1}{c|}{\makecell{25.46\\1.000}} & \multicolumn{1}{c|}{\makecell{56.82\\1.000}} & \multicolumn{1}{c|}{\textdagger} & \multicolumn{1}{c|}{\makecell{\textbf{25.06}\\1.000}} & \multicolumn{1}{c|}{\makecell{26.04\\1.000}} \\\hline
\textbf{Yelp-Review} & \multicolumn{1}{c|}{\makecell{26.75\\1.000}} & \multicolumn{1}{c|}{\makecell{56.14\\0.000}} & \multicolumn{1}{c|}{\textdagger} & \multicolumn{1}{c|}{\makecell{\textbf{25.08}\\1.000}} & \multicolumn{1}{c|}{\makecell{26.37\\1.000}} & \multicolumn{1}{c|}{\makecell{101.50\\1.000}} & \multicolumn{1}{c|}{\makecell{1607.38\\0.000}} & \multicolumn{1}{c|}{\textdagger} & \multicolumn{1}{c|}{\makecell{98.31\\1.000}} & \multicolumn{1}{c|}{\makecell{\textbf{42.04}\\1.000}} & \multicolumn{1}{c|}{\makecell{163.38\\1.000}} & \multicolumn{1}{c|}{\makecell{2881.68\\0.000}} & \multicolumn{1}{c|}{\textdagger} & \multicolumn{1}{c|}{\makecell{169.28\\1.000}} & \multicolumn{1}{c|}{\makecell{\textbf{50.15}\\1.000}} \\\hline
\textbf{Yelp-Tip} & \multicolumn{1}{c|}{\makecell{\textbf{22.37}\\1.000}} & \multicolumn{1}{c|}{\makecell{27.49\\1.000}} & \multicolumn{1}{c|}{\textdagger} & \multicolumn{1}{c|}{\makecell{22.64\\1.000}} & \multicolumn{1}{c|}{\makecell{24.38\\1.000}} & \multicolumn{1}{c|}{\makecell{29.75\\1.000}} & \multicolumn{1}{c|}{\makecell{128.89\\1.000}} & \multicolumn{1}{c|}{\textdagger} & \multicolumn{1}{c|}{\makecell{30.23\\1.000}} & \multicolumn{1}{c|}{\makecell{\textbf{25.53}\\1.000}} & \multicolumn{1}{c|}{\makecell{37.22\\1.000}} & \multicolumn{1}{c|}{\makecell{215.35\\1.000}} & \multicolumn{1}{c|}{\textdagger} & \multicolumn{1}{c|}{\makecell{40.73\\1.000}} & \multicolumn{1}{c|}{\makecell{\textbf{28.24}\\1.000}} \\\hline
\textbf{Yelp-User} & \multicolumn{1}{c|}{\makecell{\textbf{24.09}\\1.000}} & \multicolumn{1}{c|}{\makecell{49.20\\0.993}} & \multicolumn{1}{c|}{\textdagger} & \multicolumn{1}{c|}{\makecell{24.34\\1.000}} & \multicolumn{1}{c|}{\makecell{25.10\\1.000}} & \multicolumn{1}{c|}{\makecell{67.96\\1.000}} & \multicolumn{1}{c|}{\makecell{1367.74\\1.000}} & \multicolumn{1}{c|}{\textdagger} & \multicolumn{1}{c|}{\makecell{71.27\\1.000}} & \multicolumn{1}{c|}{\makecell{\textbf{34.57}\\1.000}} & \multicolumn{1}{c|}{\makecell{103.21\\1.000}} & \multicolumn{1}{c|}{\makecell{2437.43\\1.000}} & \multicolumn{1}{c|}{\textdagger} & \multicolumn{1}{c|}{\makecell{104.05\\1.000}} & \multicolumn{1}{c|}{\makecell{\textbf{40.11}\\1.000}} \\\hline
\end{tabular}
}
\\{\footnotesize \textdagger indicates the algorithm did not complete in four hours}
\caption{Runtime for distributed schema inference (runtime in seconds and fraction of accepted training documents)}\label{tbl:runtime_dist}
\end{table*}

JSONoid with the minimal set of properties has runtime similar to the approach proposed by Baazizi et al.
Performing inference for all the monoids defined by JSONoid is significantly more expensive.
However, we note that, while using the full set of monoids makes the inference process significantly slower than the fastest approach, we maintain linear scalability with respect to the number of documents.
A user is free to decide on the tradeoff between runtime and the detail provided via the chosen set of monoids.
Although not yet implemented, if detailed information is required and runtime is a concern, a sampling-based approach can be used to yield useful information from a subset of the data.

\begin{comment}
\begin{figure}
    \centering
    \includegraphics[scale=0.52]{plot.pdf}
    \caption{Runtime for distributed schema inference}\label{fig:runtime_dist}
\end{figure}
\end{comment}

As mentioned above, JSONoid is also capable of running in \emph{streaming} mode where one document is processed at a time and the schema is updated continuously.
To evaluate the performance of our various monoids in streaming mode, we consider feeding documents from the GitHub dataset used above to JSONoid one at a time and evaluating the rate of document processing.
Since we cannot directly measure the performance of an individual monoid, we start with the Minimum set of monoids and then determine the additional time taken for further monoids which are added.
We consider the overhead of each monoid added to reach the Simple set and again to reach all the monoids supported by JSONoid.
We note that since it is necessary to measure the runtime of each monoid using independent trials, summing the results of all the monoids added between the Minimum and Simple sets will not yield the exact total for each set of monoids.
Our results are given in Table~\ref{tbl:runtime_stream}.

Streaming schema discovery using the Minimum set of monoids has very high throughput.
While adding all possible monoids currently defined in JSONoid significantly reduces performance, users can pick and choose which monoids are required for their specific application.
There are also plenty of opportunities to optimize the most expensive monoids.
For example, the \textbf{Format} monoid currently serially checks multiple regular expressions against each string value.
There are several optimized regular expression engines that would allow this matching to occur in parallel~\cite{Wang2019}.

Finally, we note that while discovering all monoids across a stream has fairly low throughput, there are several additional techniques we can apply to improve performance in addition to optimizations to our particular implementation.
Firstly, we can perform sampling on the stream to reduce the need for JSONoid to keep pace with all new documents as they arrive.
Furthermore, a \emph{hybrid} mode of discovery is possible, in which the stream is partitioned into multiple streams, which are processed in parallel with each of the separately discovered schemas periodically merged.
We leave the development of this hybrid mode for future work.

\begin{table}
    \centering
    {\small
    \begin{tabular}{r|rr}
    \textbf{Monoids} & \textbf{Time (s)} & \textbf{Docs/s} \\\hline
    \multicolumn{1}{l|}{Minimum} & 59.5 & 3,681 \\\hline
    \textbf{Dependencies}        & +169.1 & \\
    \textbf{Examples}            & +455.9 & \\
    \textbf{Format}              & +310.7 & \\
    \textbf{MaxMin}              & +19.9 & \\
    \textbf{Multiple}            & +3.9 & \\
    \textbf{Pattern}             & +33.2 & \\
    \textbf{Required}            & +7.3 & \\
    \textbf{Unique}              & +5.3 & \\\hline
    \multicolumn{1}{l|}{Simple}  & 1,119.0 & 196 \\\hline
    \textbf{AttributeCounts}     & +27.2 & \\
    \textbf{Bloom}               & +197.6 & \\
    \textbf{Histogram}           & +771.5 & \\
    \textbf{HLL}                 & +636.6 & \\
    \textbf{Stats\textsuperscript{\textasteriskcentered}} & +23.4 & \\\hline
    \multicolumn{1}{l|}{All}      & 2,897.1 & 76\\\hline
    \end{tabular}
    }
    \vspace{2mm}
    \\{\scriptsize $^{*}$~The \textbf{Stats} monoid calculates statistics described in Section~\ref{subsec:stats_monoids}}
    \caption{Runtime for streaming schema inference}\label{tbl:runtime_stream}
    \vspace{-2mm}
\end{table}

\section{Use Cases}\label{sec:use_cases}

While Section~\ref{sec:eval} provided an empirical evaluation of our mining approach, here we aim to identify several use cases of the generated schemas.

\subsection{Schema Exploration}

\begin{figure}
\centering
\begin{varwidth}{\linewidth}
{\small
    \begin{verbatim}
{ "asin": "0000143561",
  "description": "3Pack DVD set...",
  "title": "Everyday Italian ...",
  "related": {
    "also_viewed": [
      "B0036FO6SI","B000KL8ODE","000014357X",
      "B0037718RC","B002I5GNVU","B000RBU4BM"],
    "buy_after_viewing": [
      "B0036FO6SI","B000KL8ODE",
      "000014357X","B0037718RC"]},
  "price": 12.99,
  "salesRank": {"Movies & TV": 376041},
  "imUrl": "http://g-ecx.images-amazon...",
  "categories": [["Movies & TV", "Movies"]]}
    \end{verbatim}
    }
\end{varwidth}
    \caption{Amazon products dataset}\label{fig:amazon_data}
\end{figure}

\begin{figure*}
    \centering
    \includegraphics[trim=0 100 100 0,clip,scale=0.5]{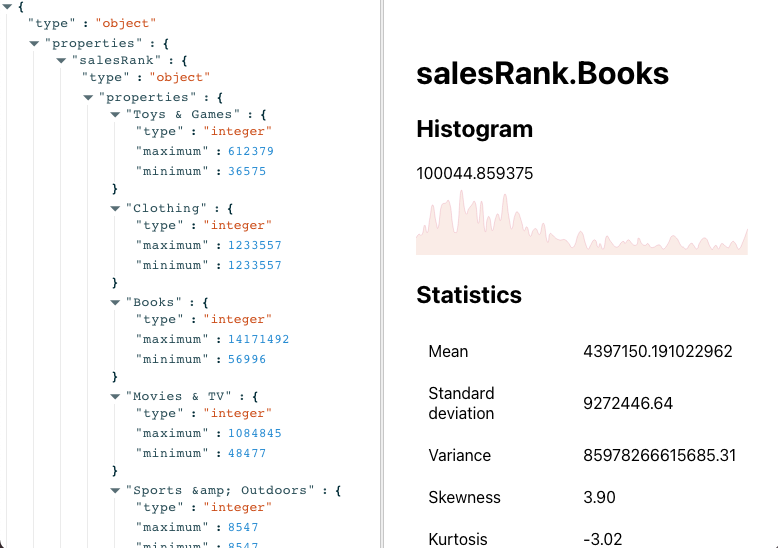}
    \caption{JSONoid screenshot with schema and some generated enhancements}
    \label{fig:screenshot}
\end{figure*}

One use of our enhancement monoids is by allowing the exploration of the generated schemas from a number of datasets along with our prescribed enhancements.
We have implemented a Web application\footnote{https://github.com/dataunitylab/jsonoid-web} that allows users to view sample documents from the collection, as well as the generated enhanced schemas and associated enhancements.
Each of the enhancements provided is applied to all applicable elements of the schema.

An example dataset is shown in Figure~\ref{fig:amazon_data} which consists of information about Amazon products~\cite{He16}.
A screenshot of the schema visualization based on these data is shown in Figure~\ref{fig:screenshot}.
On the left is the schema that was discovered based on the dataset.
On the right are some enhancements to the schema element representing the values of the key \texttt{Books} that is nested under the \texttt{salesRank} object.
In this case, the currently visible enhancements are a histogram of the observed values and statistics which were calculated across them.

These enhancements provide a detailed summary of both the structure and the values contained within the documents used to create the schema.
Without a schema, which is often unavailable for collections of JSON documents, it would be necessary for an analyst to manually browse through example documents to begin to develop an understanding of their structure and content.
This may be followed by ad hoc analyses, which provide more detailed information.
A detailed schema such as those constructed by JSONoid can alleviate much of this manual effort.

\subsection{Constraint Discovery}\label{subsec:constraints}

One important element of relational database schemas is primary and foreign key constraints.
Such constraints are typically not present in JSON schemas.
However, we make further use of monoids to estimate and suggest possible constraints.
Constraint discovery occurs as a post-processing step on the final schemas generated during the discovery process.

For attributes expected to be unique across an entire collection (i.e.\ possible primary keys), we need to track the total number of documents (a simple counter).
Furthermore, we can use HyperLogLog to estimate the total number of unique values.
When the total number of documents is with the error bounds of the number of unique values estimated by the HLL data structure constructed using the corresponding monoid, we suggest a possible primary key.
Considering our example data in Figure~\ref{fig:amazon_data}, the attribute \texttt{asin} would be suggested as a possible primary key, since the estimated count of unique values would be within the bounds of the total number of documents examined.
Although there is a small probability of false positives, it is possible to suggest this primary key without the need to maintain a list of all possible values to ensure uniqueness.
We also make use of heuristics~\cite{Souibgui2022,Papenbrock17} to provide a heuristic ranking of possible primary keys.
Although we do not further evaluate this discovery technique here, we note that the features required to compute these proposed heuristic rankings, such as data type and value length, are already captured by the schemas generated by JSONoid.

As a further proof of concept, we implemented a simplified version of the {\scshape Many}~\cite{Tschirschnitz17} inclusion dependency mining algorithm to identify possible foreign keys using the Bloom filter monoid.
Bloom filters are bit vectors, and when the bits set in one Bloom filter ($B_1)$ are a subset of the bits set in another filter ($B_2$), this indicates that the set represented by $B_1$ is likely a subset of the set represented by $B_2$.
While the Bloom filters we use are monoids which can be maintained in a distributed fashion, we do need a final pass over the collected Bloom filters to determine possible foreign keys.
We compare all pairs of Bloom filters to find possible subset relationships.
Although this operation is quadratic, it only scales relative to the size of the generated schema, rather than the size of the dataset.
In addition, the {\scshape Many} algorithm has other optimization techniques that could be incorporated in the future.
In our example in Figure~\ref{fig:amazon_data}, the values contained in the \texttt{also\_viewed} and \texttt{buy\_after\_viewing} could be suggested as possible foreign keys to the \texttt{asin} attribute.
As in the case of primary key detection, the heuristics of Papenbrock and Naumann~\cite{Papenbrock17} could prove helpful in ranking possible primary keys.
We provide a further discussion of dependency discovery on JSON data in previous work~\cite{Mior2021}.

Once these constraints are discovered, they can be used to aid in the process of normalization.
Our prior work, as well as that of DiScala and Abadi~\cite{DiScala16} has proposed approaches for generating normalized models from nested key-value data, such as JSON documents.
This can be useful for reducing redundancy in the schema and identifying possible real-world entities, making the schema more natural to work with.

\subsection{Outlier Detection}\label{subsec:outliers}

Many of the statistical monoids presented can be useful in identifying single documents that are outliers compared to the generated schema.
For example, the mean and standard deviation of collected values could determine whether particular numeric values in a document should be considered outliers.
This also applies to other numeric values that can be tracked by our monoids, such as the length of string values and arrays.

We can also compare whether schemas generated from two groups of documents are likely to have been drawn from the same underlying distribution.
That is, we can identify whether a schema created from a group of samples should be considered divergent from a schema generated from reference data.
As an example, two histograms of values can be compared using the Kolmogorov-Smirnov test~\cite{Massey51}.

Another type of outlier we can identify are \emph{structural outliers} as identified by~\cite{Klettke15} et al.
Structural outliers are attributes that occur very rarely or attributes that exist in most documents but are occasionally absent.
We are able to identify such structural outliers using our \textbf{AttributeCounts} monoid.
The result is that our schema contains the information present in Klettke et al.'s reduced structure identification graph.

All the above examples demonstrate outlier detection for a single value within a schema.
Although this may undoubtedly be useful, the question of whether \emph{documents} are outliers is more complex.
Should a single outlying value in a document result in that document being considered an outlier?
Should values present anywhere in a document be given the same weight?
While we do allow the ability to collect outlying values present in a document, we leave the question of document-level outlier detection for future work.

\section{Related Work}\label{sec:related_work}

Existing approaches to schema inference focus almost exclusively on structural and type information.
For example, Baazizi et al.~\cite{Baazizi17,Baazizi17b,Baazizi19} produce a scalable mechanism for inferring schemas.
As mentioned previously, our distributed discovery approach takes a similar approach to their first work on schema discovery, which uses Apache Spark to produce schemas for individual documents and recursively merge them~\cite{Baazizi17}.
Some of the additional information we collect includes the same information as the \emph{counting types} described in later work.
However, the resulting schema only provides information on the attributes that exist on an object and limited information on their associated types.
This is missing a rich source of useful information for analysts such as some of the examples stated in the previous subsection.
Other approaches similarly do not exploit any information within the data values themselves~\cite{Wang15,Izquierdo13}.

jHound~\cite{Moller19,Moller21} similarly provides basic structural information.
It goes one step further to identify values incorrectly represented as strings (e.g.\ \texttt{"true"} and \texttt{"false"} instead of boolean values) with the goal of presenting more useful information to data consumers.
This specific scenario was identified as common in the open data setting where jHound is applied, but it does not provide further information on the data values.
This would be a trivial addition to JSONoid via additional patterns in the $\mathbf{Format}$ monoid.

Klettke et al.~\cite{Klettke15} present an approach to schema extraction that also detects \emph{structural outliers} or documents that contain attributes that are uncommon or which do not contain attributes that are common.
The assumption is that these are either errors in the data which the consumer should be aware of or uncommon structures that the user should be prepared to handle.
Several of the metrics used to detect outliers are similar to those collected in our approach, and suggesting possible structural outliers could be a possibility for future work.
Since JSONoid is designed to process large volumes of data, we currently do not enable this use case with a single pass through the data.
However, JSONoid can provide this information on outliers in a second pass through the data based on the metadata collected.

There is also significant existing work on \emph{semantic} schema discovery~\cite{KellouMenouer2022} where the goal is to identify meaningful entities in the discovered schemas and possibly align the entities with a preexisting knowledge base.
We did not attempt to infer semantic meaning in this work, but this is an interesting area for future work.
We believe that the information extracted from JSONoid schemas could provide useful features for semantic classification.

\section{Future Work}\label{sec:future_work}

While we believe that the monoids that we present here are useful, there are many opportunities for further extension.
Specifically, we aim to consider monoids which can represent combined information from multiple attributes such as joint value distributions.
This could be helpful for tasks such as identifying possible functional dependencies.
Other information that may be useful to surface in the inference process is the identification of \emph{schema variants}.
Gallinucci et al.~\cite{Gallinucci18} explore variants within a schema such as different metadata associated with different types of log records.
Ruiz et al.~\cite{Ruiz15} and Klettke et al.~\cite{Klettke17} considered variants that arise from a schema with a structure that changes over time.
Joint statistics across attributes could also be helpful in identifying such variants.
We leave the incorporation of variant discovery as future work.

Our discussion of outlier detection in Section~\ref{subsec:outliers} raised many questions about how to identify whether a specific document is an outlier.
We may not want a single outlying value in a document to cause the entire document to be considered an outlier.
However, some combination of outlying values or structural outliers may warrant flagging a document as an outlier.
Surfacing all outlying values and structural outliers in a document and then using a supervised approach can tailor outlier detection to specific use cases~\cite{Aggarwal2017}.

To effectively use the generated schemas for tasks such as outlier detection and validation, it is necessary to avoid overfitting to the observed set of documents.
As previously discussed, we expect that for each monoid it will be possible to develop heuristics to help tune the results of these monoids to limit overfitting where that is desired.

There are several useful elements of real-world JSON schema instances which we have not touched on this work.
For example, JSON Schema allows the creation of definitions which function as an avenue for avoiding duplication within a schema.
We believe that our extracted schemas can be useful for identifying possible reuse and creating definitions in the generated schemas.
This would reduce the size of the generated schemas, as well as improve the ease of working with the schema since repeated structures would be automatically identified.
Such structures may also be useful for identifying real-world entities and normalizing denormalized collections of documents, as described previously.
We note that some previous work, such as the approach of DiScala and Abadi~\cite{DiScala16} requires materializing the entire dataset in memory, a downside we believe can be avoided using the techniques presented here.

\section{Conclusion}\label{sec:conclusion}

%\balance

We presented JSONoid, a monoid-based JSON schema discovery tool which is scalable to large collections of documents and provides significantly more useful information than existing tools.
JSONoid allows users to select a tradeoff between runtime and the information content of generated schemas.
When discovering simple information, JSONoid has performance nearly identical to those of existing approaches.
Discovering additional information has an additional cost, but the penalty remains linear in the number of documents.

We demonstrated several use cases for these schemas that are not enabled by existing schema discovery approaches.
Schemas generated by JSONoid have significantly more utility for analysts as they provide insight into the actual values contained within the documents used to create the schema.
This information enables analysts exploring the schema to make decisions such as the utility of various attributes and the sizing of data structures necessary during further processing without requiring manual browsing through documents.

For future work, we intend to explore the practical usage of JSON Schema in more detail to explore automated approaches to mine other features from collections of JSON documents that are useful for data analysts.
Two main areas where we expect JSONoid to be useful are outlier detection and the discovery of common structures within documents.

\balance

%\clearpage

%\bibliographystyle{ACM-Reference-Format}
\bibliographystyle{IEEEtran}
\bibliography{main}

\end{document}